\documentclass[prl,superscriptaddress,showpacs,longbibliography,reprint]{revtex4-1}

\usepackage{color}
\usepackage[usenames,dvipsnames]{xcolor}
\usepackage{amsmath,amsthm,amssymb}
\usepackage{graphicx}
\usepackage{epsfig}
\usepackage{dcolumn}
\usepackage{bm}
\usepackage{mathrsfs}
\usepackage{multirow}
\usepackage[all]{xy}
\usepackage{pbox}
\usepackage{verbatim}
\usepackage{braket}
\usepackage{dsfont}

\newcommand{\be}{\begin{equation}}
\newcommand{\ee}{\end{equation}}
\newcommand{\ben}{\begin{eqnarray}}
\newcommand{\een}{\end{eqnarray}}
\newcommand{\beq}{\begin{equation}}
\newcommand{\eeq}{\end{equation}}

\newcommand{\ketbra}[2]{\ket{#1}\hspace{-2.1pt}\bra{#2}}

\usepackage{graphicx}

\usepackage{color,soul}

\begin{document}

\title{Non-equilibrium phase transitions in $(1+1)$-dimensional quantum cellular automata with controllable quantum correlations}

\author{Edward Gillman}
\affiliation{School of Physics and Astronomy, University of Nottingham, Nottingham, NG7 2RD, UK}
\affiliation{Centre for the Mathematics and Theoretical Physics of Quantum Non-Equilibrium Systems,
University of Nottingham, Nottingham, NG7 2RD, UK}

\author{Federico Carollo}
\affiliation{Institut f\"{u}r Theoretische Physik, Universit\"{a}t T\"{u}bingen, Auf der Morgenstelle 14, 72076 T\"{u}bingen, Germany}

\author{Igor Lesanovsky}
\affiliation{School of Physics and Astronomy, University of Nottingham, Nottingham, NG7 2RD, UK}
\affiliation{Centre for the Mathematics and Theoretical Physics of Quantum Non-Equilibrium Systems,
University of Nottingham, Nottingham, NG7 2RD, UK}
\affiliation{Institut f\"{u}r Theoretische Physik, Universit\"{a}t T\"{u}bingen, Auf der Morgenstelle 14, 72076 T\"{u}bingen, Germany}

\begin{abstract}
Motivated by recent progress in the experimental development of quantum simulators based on
Rydberg atoms, we introduce and investigate the dynamics of a class of $(1 + 1)$-dimensional quantum cellular automata. These non-equilibrium many-body models, which are quantum generalisations of the Domany-Kinzel cellular automaton, possess two key features: they display stationary behaviour and non-equilibrium phase transitions despite being isolated systems. Moreover, they permit the controlled introduction of local quantum correlations, which allows for the impact of quantumness on the dynamics and phase transition to be assessed. We show that projected entangled pair state tensor networks permit a natural and efficient representation of the cellular automaton. Here, the degree of quantumness and complexity of the dynamics is reflected in the difficulty of contracting the tensor network. 
\end{abstract}

\date{\today}

\maketitle

\noindent {\bf \em Introduction.} Out-of-equilibrium many-body quantum systems have received considerable interest in recent years both experimentally \cite{Syassen2008,Kim2010,Barreiro2011,Bohnet2016,Gutierrez2017, Lienhard2018, Wade2018,Helmrich:2020aa} and theoretically \cite{Polkovnikov2011,Eisert2015}. Yet, a full understanding of their non-equilibrium physics remains a major challenge. In the case of quantum systems that exhibit non-equilibrium phase transitions (NEPTs), the challenge is particularly pronounced: not only are large systems required but, to study critical dynamics and steady-state behaviour, long evolution times are also needed \cite{Carollo2019,Gillman2019}.

A key question concerning NEPTs in quantum systems is the impact of quantum effects on universal physics. In classical systems, the directed percolation (DP) universality class describes the critical behaviour of a number of very different non-equilibrium models \cite{Hinrichsen2000,Henkel2008}. This includes both continuous-time models such as the classical contact process (CCP), and discrete-time models such as the $(1+1)$-dimensional stochastic Domany-Kinzel cellular automaton (DKCA). However, in the quantum regime the situation is far less clear. The quantum contact process -- a simple coherent analog of the CCP -- does not appear to belong to DP \cite{Marcuzzi2016,Buchhold2017,Roscher2018,Minjae2019,Carollo2019,Gillman2019,Jo2020}. Yet, the NEPT found previously in a $(1+1)$-dimensional quantum cellular automata (QCA) which extends the DKCA in the site-DP regime does indeed belong to DP, despite the presence of non-classical correlations \cite{Lesanovsky2019}.

In this paper, we introduce a class of $(1+1)D$ QCA that display absorbing state NEPTs and offer control over the level of quantum correlations present. This allows for the investigation of the relationship between correlations and critical dynamics. $(1+1)D$ QCA are particularly attractive as they are readily realisable in quantum simulators \cite{PhysRevLett.104.010503,Britton:2012aa,Schaus:2015aa,Labuhn:2016aa,Bernien2017,Zhang:2017aa,Banuls2019} based on Rydberg atoms \cite{PhysRevLett.104.010502,PhysRevLett.104.010503,Bloch2012,Bernien2017,Kim2018,Barredo2018,Wintermantel2019,Browaeys2020}. 
%

\begin{figure}[t]
\centering
\includegraphics[width=1\linewidth]{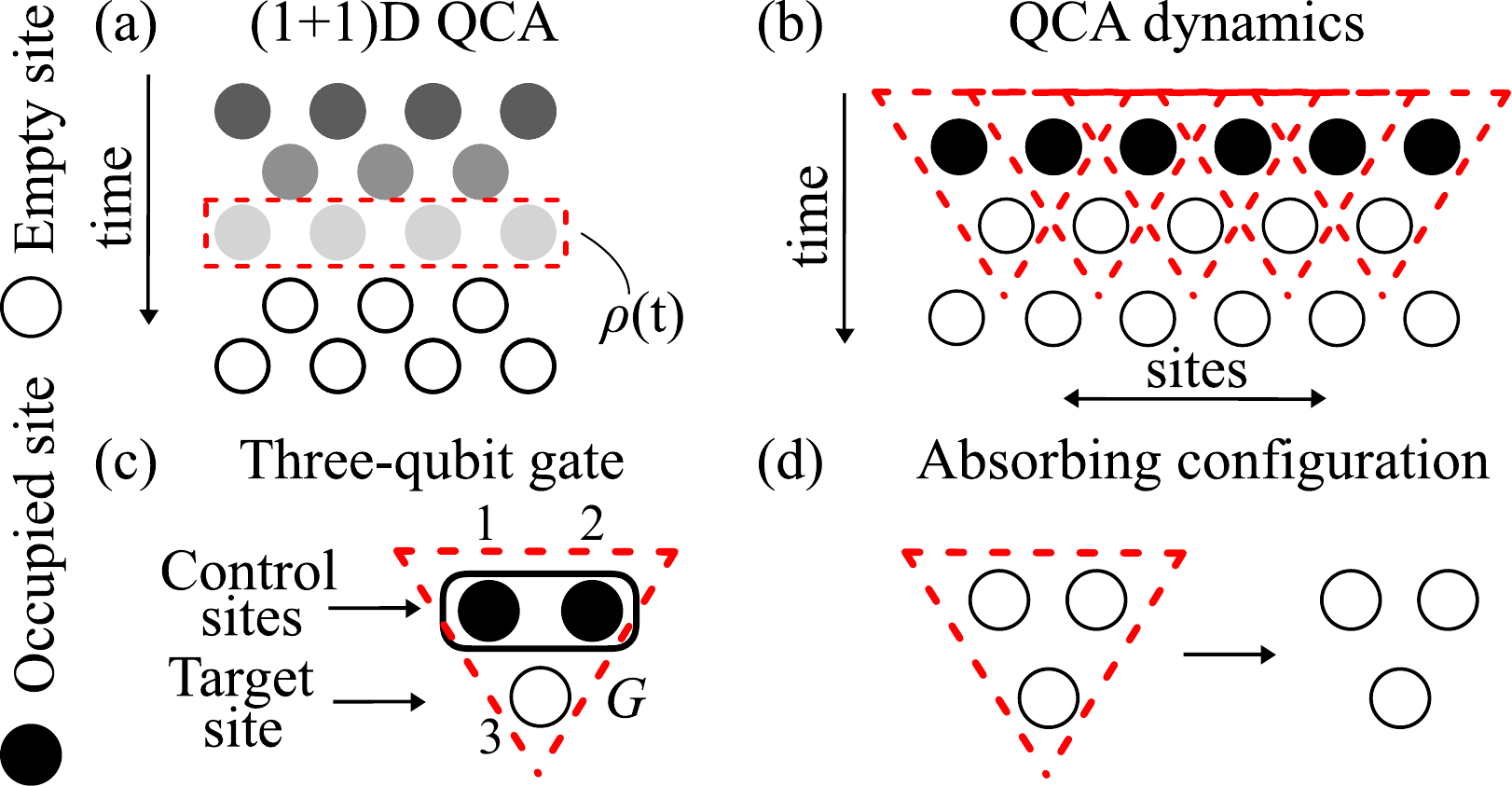}
\caption{\textbf{The $\mathbf{(1+1)D}$ QCA:} \textbf{(a)} The dynamics of the $(1+1)D$ QCA is enacted by sweeping through row-by-row with the a fundamental three-body gate, $G$. This generates an effective time-dimension in the vertical direction. The initial state is chosen to be empty at all sites apart from the first row. \textbf{(b)} $G$ is controlled by two sites, labelled by $1$ and $2$ and acts on a target one, denoted by $3$. This controlled operation defines the elementary propagation block. \textbf{(c)} Single-row observables are fully characterised by the reduced state of a row, $\rho(t)$, at the discrete time $t$. \textbf{(d)} When $G$ has an invariant configuration, the $(1+1)D$ QCA features an absorbing state.}
\label{fig:QCA_Schematic}
\end{figure}

\begin{figure*}[t!]
\centering
\includegraphics[width=1\linewidth]{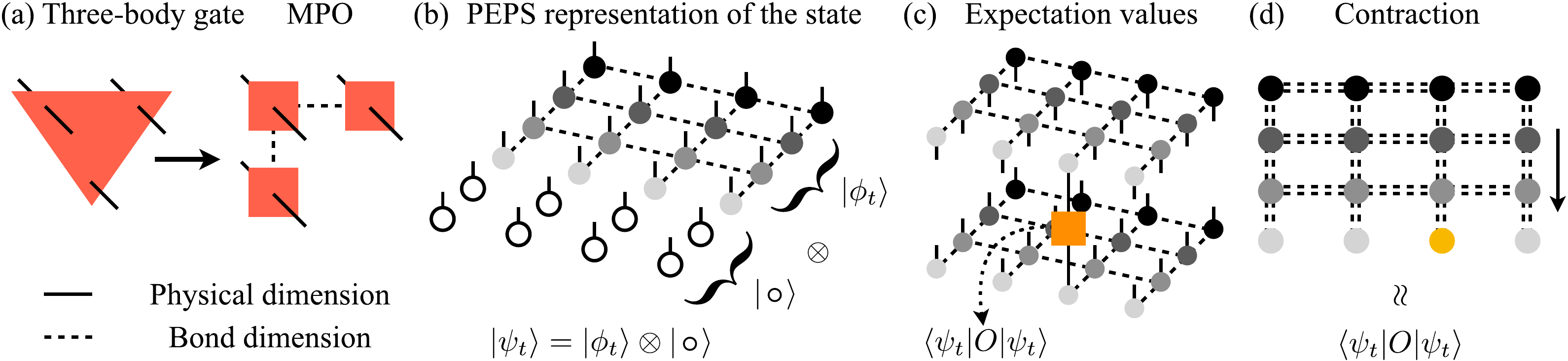}
\caption{\textbf{PEPS representation for $\mathbf{(1+1)D}$ QCA:} \textbf{(a)} The three-body gates $G$ can be written as a three-site matrix product operator (MPO) with maximum bond-dimension $4$. \textbf{(b)} The initial state is represented by an unentangled PEPS. As the MPO representation of $G$ is applied, at time-step $t$ of the evolution, each bond in the PEPS has at most bond-dimension $4$ for rows up to $t+1$, and bond-dimension $1$ otherwise. \textbf{(c)} Expectation values are represented by the double-layer TN (DLTN) obtained by sandwiching the chosen operator between the PEPS representation of the $2D$ state and its dual. \textbf{(d)} To evaluate the expectation, the DLTN must be contracted; this operation cannot be done exactly in general and requires approximation.}
\label{fig:peps_representation_construction}
\end{figure*}

Analogous to classical $(1+1)D$ cellular automata (CA) \cite{Henkel2008} a $(1+1)D$ quantum cellular automaton (QCA) consists of $2D$ lattices with sites initialized in a given state, sequentially updated, row-by-row, according to certain dynamical rules, see Fig.~\ref{fig:QCA_Schematic}. This realises an effective time-dimension in the $2D$ lattice, see Fig.~\ref{fig:QCA_Schematic}(a). To update an individual ``target" site in a row, a three body unitary gate, $G$ is applied to the site along with its nearest-neighbour parent sites, as defined by the tilted lattice to perform a controlled unitary operation, see Fig.~\ref{fig:QCA_Schematic}(b). Note that while classically $(1+1)D$ CA such as the DKCA are stochastic, here the overall evolution of the $(1+1)D$ QCA is unitary. To update an entire row, an ordering of gates must be chosen, as in general these do not commute. Following \cite{Lesanovsky2019}, we consider a sweeping pattern updating each site sequentially from left to right or vice versa. 

For $(1+1)D$ QCA we show that tensor networks (TNs), and in particular projected entangled pair states (PEPSs), provide a natural approach. We construct a TN representation of $(1+1)D$ QCA as PEPSs and outline how they can provide expectation values of local observable through a simple contraction scheme. We find that the level of local entanglement produced by $G$ significantly impacts the difficulty of contracting the PEPS, thus providing a link between quantum correlations and computational difficulty in the simulation of such QCA. We further exploit the PEPS representation to explore the quantum NEPT of the $(1+1)D$ QCA beyond mean-field (MF) techniques, and to analyse the influence of quantum effects on the corresponding transition boundary and universality class.

{\bf \em Model.} We consider a class of $(1+1)D$ QCA initially prepared with all sites in the empty state, $\ket{\circ}$, except for the first row having sites in the occupied state, $\ket{\bullet}$. We consider the gate
\begin{align}
G =\exp\left[-i \Gamma \left( P^{12} U^{12 \dagger}\sigma_-^3 +  U^{12} P^{12}  \sigma_+^3\right)\right]\, .
\label{eqn:gate_def}
\end{align}
This acts on two adjacent control sites -- labelled with $1$ and $2$ in Fig.~\ref{fig:QCA_Schematic}(b) -- via the unitary operator, $U^{12}$, and the projector, $P^{12}$, and on a target site -- denoted by the number $3$ in the Fig.~\ref{fig:QCA_Schematic}(b) -- in the subsequent row via the ladder operators, $\sigma_{+} = \ketbra{\bullet}{\circ}, \sigma_{-} = \ketbra{\circ}{\bullet}$. The projector $P^{12}$ is chosen to be orthogonal to the subspace with control sites both in the empty state, which is thus an invariant configuration for $G$, see Fig.~\ref{fig:QCA_Schematic}(d). As such, the QCA features an absorbing state: when the row $t$ consists of all empty sites, then the empty row at $t+1$ is unchanged by the update. For concreteness in what follows, we choose $P^{12} = \mathds{1} - \ket{\circ\circ}\hspace{-2.5pt}\bra{\circ\circ}$.

\begin{figure*}[t]
\centering
\includegraphics[width=1\linewidth]{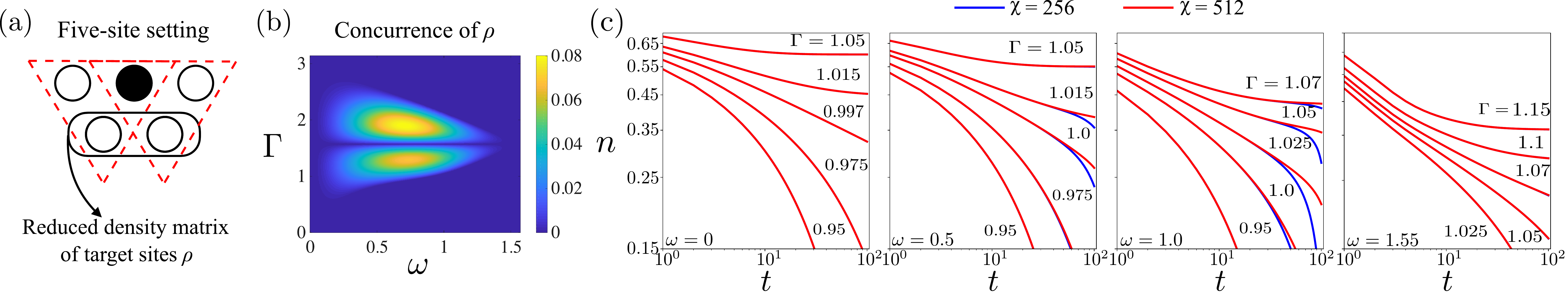}
\caption{\textbf{Gate entangling power and dynamics of the QCA.} \textbf{(a)} The entangling effect of the gate is quantified via the concurrence of the reduced density matrix of the two target sites following the action of two gates on three control sites in the state $|\!\circ\!\bullet \circ\rangle$. \textbf{(b)} Density plot of the concurrence of the reduced density matrix of the target sites. For $\omega = 0,\pi/2$ there is no entanglement, whatever the value of $\Gamma$. \textbf{(c)} Average density at the $t$-th discrete time. The approximation is obtained using the boundary MPS method with largest bond-dimension $\chi = 256$ or $512$ and up to $t=100$. We explore the range $\omega \in \left[0,\pi/2\right]$ with values of $\Gamma$ including both subcritical and supercritical curves -- identified by negative and positive curvatures in the log-log plot respectively. This reveals the NEPT of the QCA. When $\omega = 0, 1.55$, all curves for different $\chi$-values are indistinguishable on the scale shown. This contrasts the case of intermediate $\omega$ where larger $\chi$-values are required for convergence. This is broadly in line with the emergence of non-zero concurrence in panel (b).}
\label{fig:gate_concurrence}
\end{figure*}

When the two-body unitary is chosen to be identity, $G$ reduces to that studied in \cite{Lesanovsky2019}. In that case, the reduced state of a row $\rho(t)$, see Fig. \ref{fig:QCA_Schematic}(c), is separable, and can be constructed via a classical mapping to determine the universal properties of the $(1+1)D$ QCA. However, despite the existence of such a mapping, $\rho(t)$ is non-classical and features non-classical correlations in form of quantum discord \cite{Henderson:2001aa,PhysRevLett.88.017901}. Our goal is now to investigate the role of quantum entanglement, both at the level of the gate application and in $(1+1)D$ QCA as a whole. To this end we introduce the two-body unitary,
\begin{align}
U^{12}\equiv U^{12}(\omega) = \exp\left(-i\omega\left[\sigma_z^1 \sigma_y^2 +\sigma_y^1 \sigma_z^2\right]\right) ~ ,
\end{align}
where $\sigma_{y} = -i \ketbra{\bullet}{\circ} + i \ketbra{\circ}{\bullet}, \sigma_{z} = \ketbra{\bullet}{\bullet} - \ketbra{\circ}{\circ}$ are Pauli spin operators, acting on the control sites $1$ and $2$. $U^{12}$ is capable of entangling adjacent control sites and thereby also permits the parametric control of the entanglement between target sites via $\omega$ [see discussion further below and also Fig. \ref{fig:gate_concurrence}(a)]. When $\omega = 0$,  $U^{12} = \mathds{1}$ and the gate reduces to that of Ref.~\cite{Lesanovsky2019}, where $\rho(t)$ is separable. However, for $\omega\neq0$, this is no longer the case, allowing for the investigation of the effects of quantum correlations on its dynamics.

Note that the term QCA is also used in the quantum information (QI) context \cite{Brennen2003,Schumacher2004,Wiesner2009,Gutschow:2010aa,Cirac2017,Gopalakrishnan_2018,Arrighi2019,Farrelly2019}, e.g. to denote certain computational models. The update $\rho(t) \to \rho(t+1)$ can be viewed as a 1D open QCA in the QI sense, when such discrete-time dynamics feature a strict light-cone structure, achieved, e.g., via appropriate orderings of the gates in the (1 + 1)D QCA \cite{Arrighi2019,Farrelly2019}. However, $\rho(t)$ can only be used to calculate the subset of $(1+1)D$ QCA observables that have support on a single row. On the other hand, observables such as unequal ``time" correlation functions require the state of the full $2D$ lattice. 

{\bf \em PEPS representation.} For large-scale simulations, the $(1+1)D$ QCA lends itself naturally to a representation in terms of PEPSs \cite{Verstraete2004,Orus2019}. Fixing a product many-body basis for the $2D$ lattice with elements $\ket{\mathbf{s}}$, a PEPS representation of a state $\ket{\psi}$ expresses its overlap with a basis state, $\psi(\mathbf{s})\equiv \braket{\mathbf{s}|\psi}$, as the contraction of a set of tensors $A^{[i,j]}$, one per lattice site. In the bulk, tensors are of rank $5$. One leg of each tensor is associated with the physical site index, while the remaining ones are virtual legs encoding correlations in the state. The dimensions of these legs are known as the bond-dimensions of the PEPS and we denote the largest as $\chi$. The bond-dimension is closely related to the maximum amount of entanglement that can be supported by the PEPS \cite{Eisert2013} and determines the cost of the representation: typically, the higher the entanglement in $\ket{\psi}$, the higher the computational costs. In the set up we consider, the initial state is unentangled. Entanglement is only built up when gates are applied, with each gate $G$ acting on a non-extensive number of sites. This allows for an efficient PEPS representation of the QCA: indeed, because of this, the full $2D$ state always satisfies an entanglement area-law.

To construct the PEPS, we begin by decomposing the gate $G$ into a three-site matrix product operator (MPO), see Fig.~\ref{fig:peps_representation_construction}(a). When applied to the initial state, a PEPS with $\chi = 1$, the new state is also a PEPS but with $\chi \le 4$. During the time-evolution each bond is acted on only once, resulting in a PEPS with again $\chi \le 4$. From Fig.~\ref{fig:peps_representation_construction}(b) the structure of the state at the discrete time $t$ is apparent: the non-trivial part, $\ket{\phi_t}$, is supported on the first $t+1$ rows, where gates have been applied.

With the PEPS representation, any desired observables can be represented also as a TN. For expectation values this is formed by sandwiching a chosen operator between the PEPS state and its dual to form a double-layer TN (DLTN), see Fig.~\ref{fig:peps_representation_construction}(c). In the $(1+1)D$ QCA considered, the state is represented exactly as a PEPS at any time and, remarkably, this means that an exact DLTN representation for expectation values is possible, in contrast to  common scenarios \cite{Orus2019}. However, the calculation of an expectation requires the contraction of the DLTN, see Fig.~\ref{fig:peps_representation_construction}(d). This cannot be done efficiently and one must resort to approximation schemes \cite{Schuch2007,Schwarz2016,Haferkamp2018,Lubasch2014,Orus2019}. Here, we consider the simplest method based on the construction of boundary matrix product states (MPSs) \cite{Verstraete2004,Orus2019}. Namely, one considers the first row of the DLTN as an MPS, typically used to represent one-dimensional quantum systems \cite{Schollwock2011,Wall2012,Montangero2018,Silvi2019}. Subsequent rows are then viewed as MPOs that enact an effective time-evolution on the MPS. This non-unitary ``time-evolution" is approximated using standard methods for MPS \cite{Schollwock2011,Jaschke2018, Paeckel2019}. The computational costs then depend on the bond-dimension required for the boundary MPS to accurately reproduce the desired expectation. In the $(1+1)D$ QCA, the boundary MPS before the final contraction represents the ``past": from this viewpoint, the bond-dimension required is a measure of the boundary MPS entanglement, encoding the correlations built by the dynamics. We note that such entanglement is not the physical entanglement of the QCA, but it rather gives a notion of the difficulty of the simulation.

{\bf \em Dynamical simulations and quantum correlations.} To investigate the impact of entanglement on the $(1+1)D$ QCA dynamics, and to link it to the difficulty of performing computations with PEPS, we first consider the possibility of generating entanglement through the gate $G$. To this aim, we consider a five-site setting with three adjacent control sites and two targets, see Fig.~\ref{fig:gate_concurrence}(a). We take the first and the third control sites to be empty, while the second is occupied. As a function of the gate parameters, $\Gamma$ and $\omega$, we calculate the concurrence \cite{PhysRevLett.78.5022} -- a measure of entanglement -- of the reduced density matrix of the two target sites. As shown in Fig.~\ref{fig:gate_concurrence}(b), when $\omega = 0$ the concurrence is zero for all values of $\Gamma$. The state $\rho(t)$ is separable though non-classical correlations are still present \cite{Lesanovsky2019}. In contrast, away from this point, entanglement is generated: this initially increases with $\omega$ but then vanishes again as $\omega\to\pi/2$.

With this in mind, we investigate the absorbing state NEPT in our many-body $(1+1)D$ QCA. Namely, we follow the evolution of the average density, $n(t)$, defined as,
\begin{equation}
n(t) = \frac{1}{N}\sum_{j=1}^{N}{\rm Tr}\left(\rho(t) \, \hat{n}_{j}\right)\, ;
\end{equation}
$\hat{n}_{j}=\ketbra{\bullet}{\bullet}_j$ is the number operator for the $j$-th site of the QCA, and $N$ is the number of columns in the $2D$ lattice. To minimise boundary effects, we find it sufficient, for our range of parameters, to fix $N = 128$ while also alternating left/right sweeps in the application of $G$.

Fig.~\ref{fig:gate_concurrence}(c) shows the evolution of $n(t)$ for boundary MPS with $\chi = 256$ or $512$. In the first plot, $\omega=0$, curves for different $\chi$ overlap, meaning that the estimated value of $n(t)$ has converged. We notice that $n(t)$ displays an emergent critical dynamics reminiscent of a second-order absorbing state NEPT \cite{Henkel2008}. Moreover, the critical value that we estimate, $\Gamma_{\rm c}\approx 0.997$, and the power-law behavior, $n(t)=t^{-\alpha}$, with $\alpha = 0.157$ - obtained via a power-law fit for $t \in [30,100]$ to the corresponding curve - is in close agreement with the exactly known ones (see Ref.~\cite{Lesanovsky2019}).

From Fig.~\ref{fig:gate_concurrence}(c), we can see that the structure of the NEPT persists for $\omega > 0$. However, the numerical simulations also become less accurate for a fixed bond dimension - and thus more costly - with $\chi = 256$ differing significantly from $\chi = 512$ when $\omega = 0.5, 1.0$. When $\omega \approx \pi/2$ the difficulty of the simulations reduces again, as shown by the fact that the agreement between the different $\chi$-curves improves. Despite their different meaning, the behaviour of the required $\chi$ (which characterises the strength of the correlations in the boundary MPS) as a function of $\omega$ shows the same features as that of the entanglement generated by the gate. In particular, we notice how it is really the capability of the gate to generate entanglement which makes simulations difficult, while the presence of other quantum correlations, such as discord (present already for $\omega=0$ \cite{Lesanovsky2019}), does not seem to play a crucial role in the accuracy of the simulations.

\begin{figure}[t]
\centering
\includegraphics[width=0.8\linewidth]{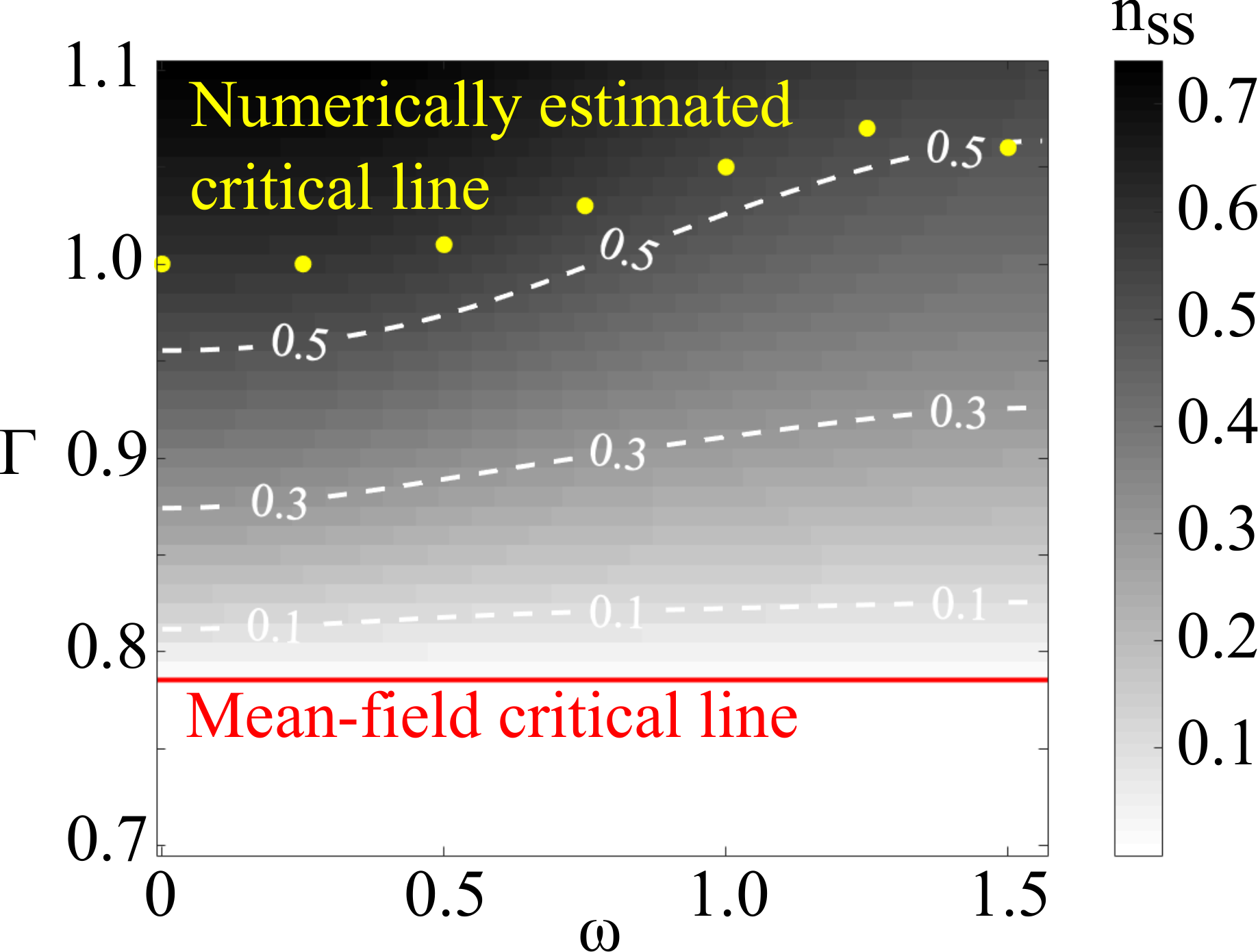}
\caption{\textbf{Phase Diagram of the $\mathbf{(1+1)D}$ QCA}. The stationary state density, $n_{\rm ss}$, estimated using a (mean-field) MF approach, is shown as a contour-plot with the indicated colour map. The corresponding MF phase boundary is displayed as a red solid line, and shows no dependence on $\omega$. By contrast, the phase boundary extracted from PEPS simulations (yellow circles) shows that the critical $\Gamma$ does indeed depend on the parameter $\omega$ that controls the local entanglement.}
\label{fig:phase_diagram}
\end{figure}

{\bf \em Phase diagram and comparison with mean-field.} To explore the possibility of our method to characterize critical behaviour beyond MF, and any potential impact of quantum effects on such behaviour, we first consider a MF approach to construct an estimate of the phase diagram for the model under investigation. The MF method, adapted from Ref.~\cite{Lesanovsky2019} and detailed in the supplemental material \cite{SM}, considers a product ansatz for $\rho(t)$ and exploits a five-site setting, {\it c.f.}~Fig.~\ref{fig:gate_concurrence}(a), to perform a time update the reduced state. 

The MF phase boundary is indicated by a solid red line in Fig. \ref{fig:phase_diagram} and shows no dependence on $\omega$. This contrasts the results of PEPS simulations, shown in Fig. \ref{fig:phase_diagram} by yellow circles. In that case, critical values of $(\Gamma, \omega)$ were estimated via power-law fits over $t \in [20,50]$ with $\chi = 256$ simulations. Taking a set of $\Gamma = 1,1.01,1.02,...,1.1$ for each $\omega$, the $\Gamma$ that was best fit by a power-law was selected as the estimated critical value. The resulting phase boundary shows a clear dependence on $\omega$. Interestingly, its shape is similar to the lines of constant density obtained from MF, which indicates that our MF approach might become reliable away from criticality.

To further test the PEPS method, we characterize the universality class of the NEPT by considering the value of a key critical exponent, $\alpha$, associated to the power-law decay of $n(t) \sim t^{-\alpha}$ at criticality \cite{Hinrichsen2000}. When $\omega = 0$, the universality class of the model is known to be that of $1D$ DP \cite{Lesanovsky2019}. This class has been studied extensively in classical theories \cite{Henkel2008}, and the exponent $\alpha \approx 0.16$, in contrast to the MF DP result $\alpha_{\text{MF}} = 1$. Given the robustness of the DP universality, one might expect this to persist also when $\omega > 0$.

Applying the MF method, we find an estimated value $\alpha \approx 1$ for all considered $\omega$. Due to the low computational costs of the MF method, this can be achieved easily via power-law fits to the resulting $n(t)$ curves. In the case of PEPS simulations, rather than estimate $\alpha$ directly, it is easier to establish bounds on its value. This can be achieved by power-law fits to super and subcritical curves \cite{Henkel2008,Carollo2019}, as detailed in the supplemental material \cite{SM}. We find that for all values of $\omega$ tested the bounds found are consistent with $1D$ DP. Furthermore, other potentially relevant classes such as MF DP, $2D$ DP or that of the quantum contact process \cite{Carollo2019,Gillman2019} can be ruled out, leading us to the conclusion that $1D$ DP is indeed the universality class of the model for all $\omega$-values. 

{\bf \em Conclusions and Outlook.}
Our work provides a basis for several directions of future investigation. Firstly, while the universal properties at the phase transition here appear to be unaffected by entanglement, this contrasts absorbing-state NEPTs in similar (continuous) open quantum systems \cite{Marcuzzi2016,Buchhold2017,Roscher2018,Minjae2019,Carollo2019,Gillman2019}. The source of this difference, and the impact of quantum effects, is an important question for further understanding of NEPTs. Secondly, linking our $(1+1)D$ QCA and those used in QI -- e.g. by understanding the role of $\rho(t)$ as a potential $1D$ open QCA -- may extend the regime of applicability of our TN method to models of interest in QI. Finally, the methods can be used to make quantitative predictions for experimental realisations of quantum NEPTs in quantum simulators \cite{Lesanovsky2019}. In turn, experiments also provide highly non-trivial tests for TNs, e.g. for the accuracy of different contraction schemes  -- the development of which is currently a significant area of research.

\begin{acknowledgments}

\acknowledgments
{\bf \em Acknowledgements.} We acknowledge support from EPSRC [Grant No. EP/R04421X/1], from the ``Wissenschaftler R\"{u}ckkehrprogramm GSO/CZS" of the Carl-Zeiss-Stiftung and the German Scholars Organization e.V., as well as through The Leverhulme Trust [Grant No. RPG-2018-181], and the DPG SPP 1929 (GiRyd). F.C. acknowledges support through a Teach@T\"{u}bingen Fellowship. We are grateful for access to the University of Nottingham's Augusta HPC service. We acknowledge the use of Athena at HPC Midlands+, which was funded by the EPSRC on grant EP/P020232/1, in this research, as part of the HPC Midlands+ consortium.

\end{acknowledgments}


\bibliographystyle{apsrev4-1}
\bibliography{DTQCA}

\onecolumngrid
\newpage

\renewcommand\thesection{S\arabic{section}}
\renewcommand\theequation{S\arabic{equation}}
\renewcommand\thefigure{S\arabic{figure}}
\setcounter{equation}{0}
\setcounter{figure}{0}

\begin{center}
\textbf{\large Supplemental Material}
\end{center}
\vspace{0.1cm}
\begin{center}
\textbf{\large Critical Quantum Dynamics in $\mathbf{(1+1)}$-dimensional Quantum Cellular Automata with Projected Entangled Pair States}
\end{center}
\begin{center}
Edward Gillman$^{1,2}$, Federico Carollo,$^{1,2,3}$ and Igor Lesanovsky$^{1,2,3}$
\end{center}
\begin{center}
$^1${\em School of Physics and Astronomy, University of Nottingham, Nottingham, NG7 2RD, United Kingdom}\\
$^2${\em Centre for the Mathematics and Theoretical Physics of Quantum Non-Equilibrium Systems, University of Nottingham, Nottingham, NG7 2RD, UK, United Kingdom}\\
$^3${\em Institut f\"{u}r Theoretische Physik, Universit\"{a}t T\"{u}bingen, Auf der Morgenstelle 14, 72076 T\"{u}bingen, Germany}
\end{center}

\section*{Details on the mean-field dynamics of the quantum cellular automata }
\label{supp:MF_Approach}

In this section we discuss in more detail the uncorrelated ansatz solution to the reduced dynamics of the quantum cellular automata that is presented in the main text. The main assumption is that the reduced state of the system is considered to be in a product state at each time-step
\begin{equation}
\rho(t)=\varphi_t^1 \varphi_t^2\dots \varphi_t^N\, ,
\end{equation}
where each site of the $t$-th row is described by the same single-site density matrix $\varphi_t$. To obtain an approximation for the time-evolution of $\rho(t)$, it is thus sufficient to determine how, through the gates $G$, the matrix $\varphi_t$ is updated.

\begin{figure}[b]
\centering
\includegraphics[width=0.6\linewidth]{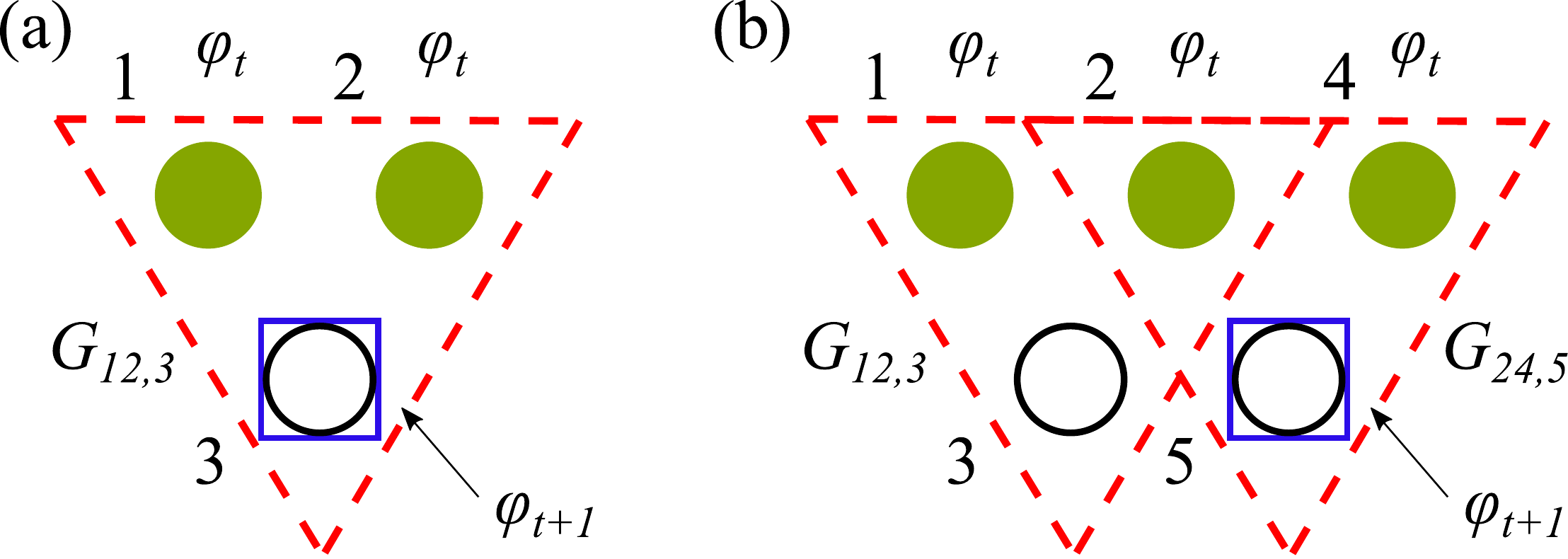}
\caption{\textbf{Mean-field updates of the quantum cellular automata}. Two different approximate update schemes for the single-site density matrix $\varphi_t$. (a) At the $t$-th time step, within the mean-field approximation, two control sites are described by the single-site density matrix $\varphi_t$. The application of the gate involves also a target site. By tracing over control sites, one then obtains a density-matrix which is the one associated with the target site. This matrix, which we call $\varphi_{t+1}$, can be taken to define the reduced state of the quantum cellular automata at the next time-step $\rho(t+1)$. (b) The previous scheme can be augmented by considering a five-site setting. In this case, we have three control sites and two target ones. At time $t$, all control sites are in the state $\varphi_t$; the updated matrix $\varphi_{t+1}$ is obtained by tracing out all sites but the fifth one, after the application of two gates. }
\label{fig:mean-field-SM}
\end{figure}

In Ref.~\cite{Lesanovsky2019}, this was achieved by considering a single plaquette formed by two control sites and a single target. As shown in Fig.~\ref{fig:mean-field-SM}(a), defining the gate $G_{12,3}$ as the gate $G$ acting on control sites $1$ and $2$ and on the target site $3$, the updated matrix $\varphi_{t+1}$, in this scheme, is obtained as
$$
\varphi_{t+1}={\rm Tr}_{12}\left(G_{12,3}\, \Big[\varphi_t^1\varphi_t^2 (\ket{\circ}\bra{\circ})^3\Big]\, G_{12,3}^\dagger\right)\, ,
$$
where ${\rm Tr}_{12}$ is the trace over the degrees of freedom of the two control sites. While this scheme is attractive due to its simplicity, it does not lead to any dependence on the parameter $\omega$ characterizing the gate $G$ in the quantum cellular automaton we consider.

As such, in the main text we adopt an improved version of the update scheme considering a five-site setting. The control sites $1,2,4$ are in a product state with single-site density matrix $\varphi_t$, while target sites $3,5$ are in the empty state. To define the matrix $\varphi_{t+1}$ of the next time-step we apply the gates $G_{12,3}$ and then $G_{24,5}$. Tracing over all sites but the fifth one gives the $\varphi_{t+1}$. This update scheme is sketched in Fig.~\ref{fig:mean-field-SM}(b), and is  given by the following iterative equation
$$
\varphi_{t+1}={\rm Tr}_{1234}\left(G_{24,5}G_{12,3}\, \Big[\varphi_t^1\varphi_t^2 (\ket{\circ}\bra{\circ})^3\varphi_t^4 (\ket{\circ}\bra{\circ})^5\Big]\, G_{12,3}^\dagger G_{24,5}^\dagger\right)\, .
$$

\section{Estimation of Critical Exponents for $(1+1)D$ quantum cellular automata}
\label{supp:Exponents_Est}

In this section, we discuss the estimation of critical exponents outlined in the main text. In absorbing state phase transitions, time-dependent order parameters can display universal dynamical scaling \cite{Henkel2008}. In particular, starting from a homogeneous, fully-occupied initial state, at criticality the average density, $n(t)$ displays a power-law decay,
\begin{align}
n(t) \sim t^{-\alpha} ~,
\end{align}
where $\alpha$ is the associated critical exponent. In the case of directed percolation (DP), the value of $\alpha$ has been established through extensive numerical simulations as $\alpha = 0.159, 0.451$ and $0.732$  for one, two and three-dimensions respectively. In four-dimensions and above, the mean-field (MF) value, $\alpha_{\text{MF}} = 1$, is valid.

To estimate of the value of $\alpha$, one can bound it from below (above) by considering super-critical (sub-critical) curves of $n(t)$ \cite{Hinrichsen2000}: After an initial transient period, curves close to criticality will display a power-law behaviour up to a time set by the temporal correlation length, $\xi_{\parallel}$, which diverges at the critical point. After this, such curves will deviate via an exponential decay to either zero, for sub-critical curves, or a non-zero constant, for super-critical curves. As such, in a log-log plot, super-critical curves show positive curvature while sub-critical curves display negative curvature.

By fitting power-laws to super (sub) critical curves, one can thus estimate lower (upper) bounds on the value of $\alpha$. When dealing with quantum many body systems, this method presents a significant advantage over direct estimates of $\alpha$ via a search for the critical point and subsequent fit to the critical line: As one only needs to obtain $n(t)$ up to a time where there is clear deviation from the power-law (in order to distinguish whether the curve is sub or super critical) shorter evolution times are required to establish the bounds than direct estimates. Of course, to get tighter bounds one must consider curves closer to criticality, which then require longer evolution times to distinguish. However, in situations where long evolution times present a significant challenge, as in quantum many body systems, this processes nonetheless allows one to get bounds on $\alpha$ that are reliable and sufficient to eliminate potential universality classes.

\begin{figure}[b]
\centering
\includegraphics[width=0.8\linewidth]{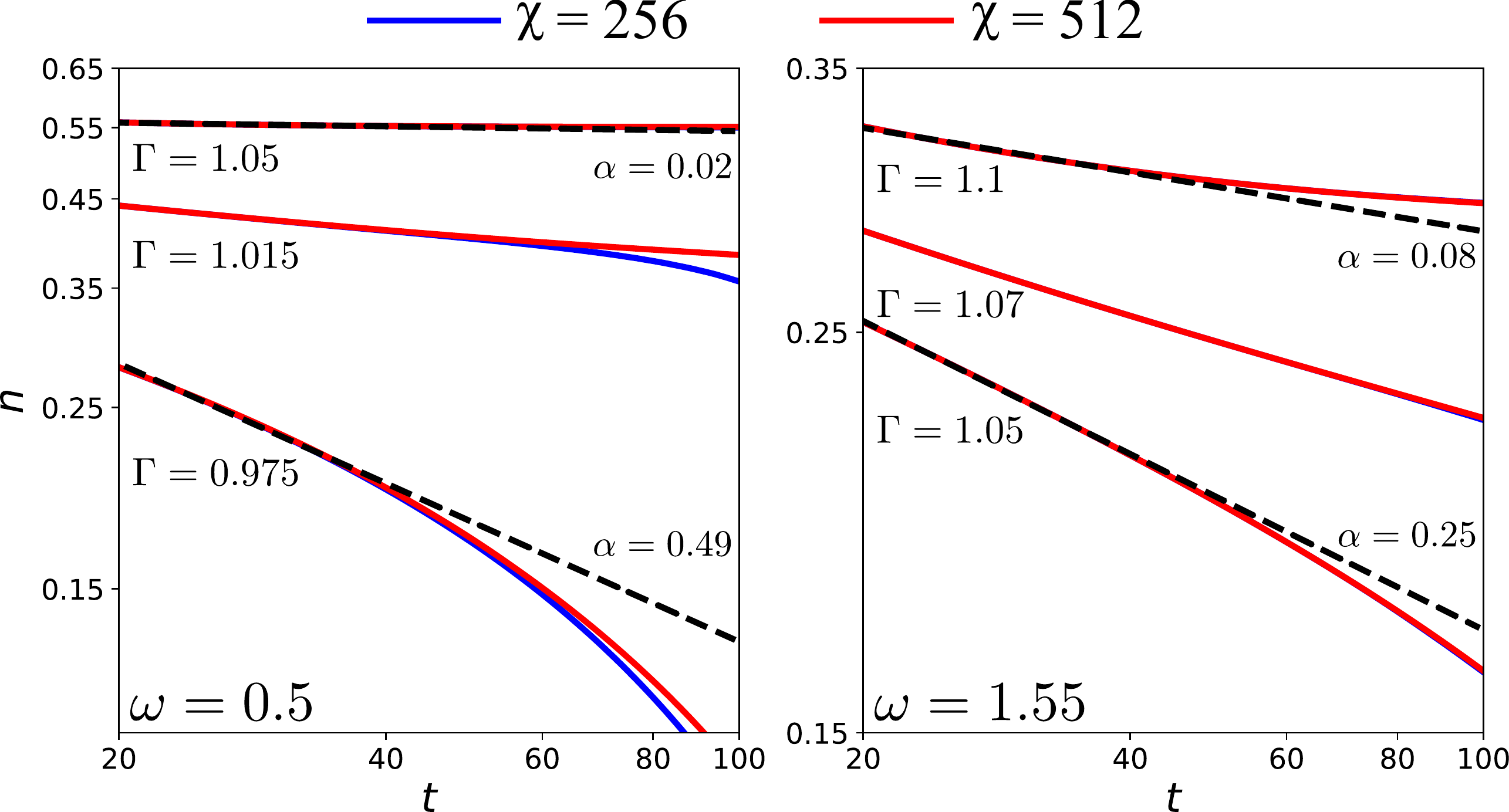}
\caption{\textbf{Bounds on universal exponent}. By using converged super and sub critical curves, bounds on the $\alpha$ exponent can be obtained for each $\omega$. Given the difference in computational difficulty for simulations with different $\omega$ -- that is, the value of $\chi$ required for convergence -- for some regions of the phase diagram significantly tighter bounds can be found than for others. This is illustrated for $\omega = 0.5$ and $\omega = 1.55$, representative of relatively challenging and simple regions of the phase diagram respectively. As can be seen, in the case of $\omega = 0.5$, convergence is reached only for curves relatively far from the critical point, when compared to $\omega = 1.55$. When making power-law fits to extract the exponent bounds, here performed for $t \in \left[20, 40\right]$, this leads to the significantly looser bounds for $\omega = 0.5$ of $\alpha \in \left[0.02, 0.49\right]$ compared to $\alpha \in \left[0.08, 0.25\right]$ for $\omega = 1.55$.}
\label{fig:exponent_bounds}
\end{figure}

Fig. \ref{fig:exponent_bounds} illustrates this for the cases of $\omega = 0.5$ and $\omega = 1.55$. Simulations are performed with $\chi = 256, 512$. For $\omega = 1.55$, the different bond-dimension curves overlap closely on the scale shown. In contrast, for $\omega = 0.5$, significant differences can be observed, corresponding to a higher required computational costs for convergence up to a given time. As a consequence, one is able to study curves closer to criticality for $\omega = 1.55$ than for $\omega = 0.5$. This leads to tighter bounds on the value of $\alpha$, which can be obtained by performing power-law fits. These were performed for $t \in \left[20, 40\right]$ to the $\chi = 512$ curves, leading to the bounds $\alpha \in \left[0.02, 0.49\right]$ and $\alpha \in \left[0.08, 0.25\right]$ for $\omega = 0.5, 1.55$. Similar bounds can be obtained for other values of $\omega$, and we additionally check $\omega = 0, 1.0$. In every case we find bounds that are consistent with $1D$ DP, and inconsistent with the mean-field value.

In addition to mean-field, for some values of $\omega$ other universality classes can also be eliminated. This includes potentially relevant classes such as $2D$ DP, $3D$ DP, and those associated with the quantum contact process \cite{Carollo2019,Gillman2019}, which are all inconsistent with the bounds obtained for $\omega = 1.55$. However, in more challenging regions of the phase diagram, such as when $\omega = 0.5$, the established bounds are wide enough to exclude only some of these, e.g., $3D$ DP. Nonetheless, there is no evidence to support the idea that the universality class changes with $\omega$, and we conclude that $1D$ DP is indeed the universality class of the model under consideration.

\bibliographystyle{apsrev4-1}
\bibliography{DTQCA}

\end{document}